%% file: main.tex
\documentclass[12pt]{article}
\usepackage{times}
\usepackage{geometry}
\usepackage[utf8]{inputenc}
\usepackage{enumitem,amssymb}
\usepackage{ragged2e}
\usepackage{graphicx}
\usepackage{sidecap}
\usepackage[super,sort&compress]{natbib}

\newlist{thematic}{itemize}{8}
\setlist[thematic]{label=$\square$}
\usepackage{pifont}
\newcommand{\cmark}{\ding{51}}%
\newcommand{\done}{\rlap{$\square$}{\raisebox{2pt}{\large\hspace{1pt}\cmark}}%
\hspace{-2.5pt}}

\usepackage[superscript,biblabel]{cite}

\usepackage{aas_macros}

\usepackage{xcolor}
\usepackage[font=small,labelfont=bf]{caption}

\newcommand{\Gaia}{\emph{Gaia}}

\begin{document}
\raggedright
\huge
Astro2020 Science White Paper \linebreak

The Multidimensional Milky Way
\linebreak
\normalsize

\noindent \textbf{Thematic Areas:} \hspace*{60pt} $\square$ Planetary Systems \hspace*{10pt} $\square$ Star and Planet Formation \hspace*{20pt}\linebreak
$\square$ Formation and Evolution of Compact Objects \hspace*{31pt} \done Cosmology and Fundamental Physics \linebreak
  \done  Stars and Stellar Evolution \hspace*{1pt} \done Resolved Stellar Populations and their Environments \hspace*{40pt} \linebreak
  \done    Galaxy Evolution   \hspace*{45pt} $\square$             Multi-Messenger Astronomy and Astrophysics \hspace*{65pt} \linebreak
  
\textbf{Principal Author:}

Name: Robyn Sanderson	
 \linebreak						
Institution:  University of Pennsylvania / Flatiron Institute
 \linebreak
Email: robynes@sas.upenn.edu
 \linebreak
% Phone:  (Seriously??)
%  \linebreak
 
%\textbf{Co-authors:} Ting, Chervin, Nico, Emily, Tony, Jeff, Kathryn. Members of WFIRST Astrometry WG, MSE science team, others
\input{authors.tex}

%  \linebreak
  
\justify
\textbf{Abstract: 
Studying our Galaxy, the Milky Way (MW), gives us a close-up view of the interplay between cosmology, dark matter, and galaxy formation. In the next decade our understanding of the MW's dynamics, stellar populations, and structure will undergo a revolution thanks to planned and proposed astrometric, spectroscopic and photometric surveys, building on recent advances by the \emph{Gaia} astrometric survey. Together, these new efforts will measure three-dimensional positions and velocities and numerous chemical abundances for stars to the MW's edge and well into the Local Group, leading to a complete multidimensional view of our Galaxy. Studies of the multidimensional Milky Way beyond the \emph{Gaia} frontier---from the edge of the Galactic disk to the edge of our Galaxy's dark matter halo---will unlock new scientific advances across astrophysics, from constraints on dark matter to insights into galaxy formation.}

\pagebreak
% Insert your white paper text here (max of five pages including figures).

%\section{Introduction}

\noindent\fbox{%
    \parbox{\textwidth}{
\vspace{-5mm}
\section*{Executive summary}
\vspace{-3mm}
%\begin{itemize}
    $\bullet$ Mapping the dark matter distribution in the outer Milky Way by dynamically modeling the measured positions and velocities of stars, globular clusters, and dwarf galaxies will enable us to {\bf constrain dark matter models} by placing our Galaxy into its cosmological context.
    
    \vspace{1mm}
     $\bullet$ Mapping dynamical and chemical variations of stars in the outer halo will {\bf establish the Galaxy's accretion history} and allow us to {\bf study galaxy formation across cosmic time, over six orders of magnitude in stellar mass}.    
     
    \vspace{1mm}
     $\bullet$ Different stellar populations used to map the Galaxy offer a {\bf tradeoff between density and volume}. These {\bf tracers must be combined} in order to leverage their different advantages, especially those offered by {\bf standard candles} such as RR Lyrae stars. {\bf Reaching the main sequence turnoff will be crucial} to place meaningful constraints on the Galactic accretion history and the response of the Galactic disk and stellar halo to perturbations from mergers.
         
    \vspace{1mm}
     $\bullet$ Creating this powerful new view of our Galaxy will require synthesizing {\bf photometry} from LSST, {\bf astrometry} from \emph{Gaia}, LSST, and WFIRST, and {\bf spectroscopy} from 4, 10, and 30-meter-class ground-based telescopes. {\bf Efficient, flexible, open-source} strategies for cross-matching, joint analysis, and synthetic observations will be crucial to this effort.
         
    \vspace{1mm}
     $\bullet$ The groundbreaking scale and dimensionality of this new view of the Galaxy challenges us to develop {\bf new theoretical tools to robustly compare these data to simulations}, and {\bf new, flexible models for the Milky Way's mass distribution}. Numerical experiments, constrained N-body models and fully cosmological-hydrodynamical simulations will all play important roles, but require {\bf continued access to computational resources beyond a focus on the ``cutting edge.''}
%\end{itemize}
%\vspace{-5mm}
}
}
\vspace{-7mm}

\section*{Science in the Multidimensional Milky Way}
\begin{figure*}
\centering
    \begin{tabular}{cc} \hspace{-3mm}
     \includegraphics[height=2.85in]{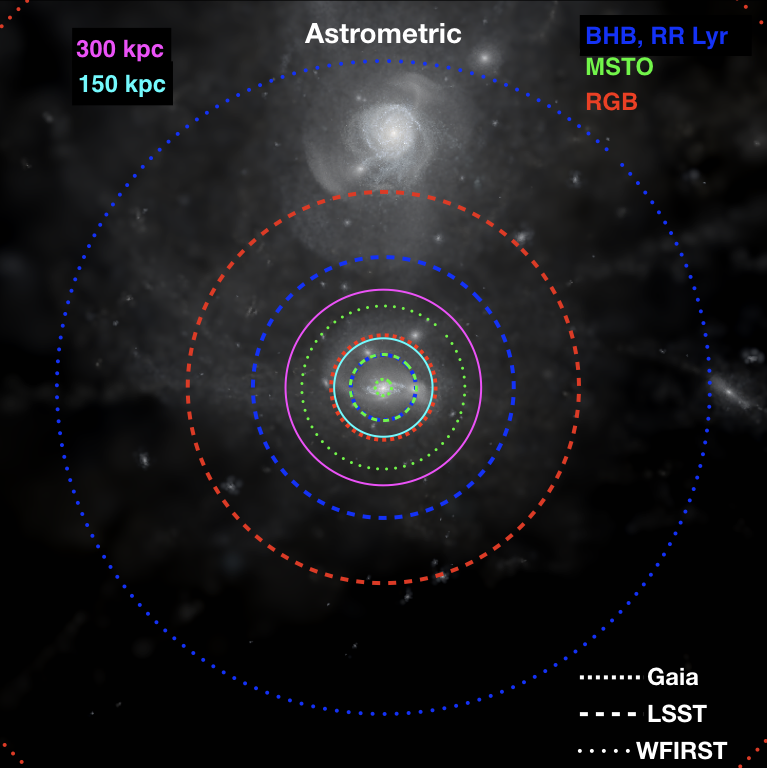} &
    \includegraphics[height=2.85in]{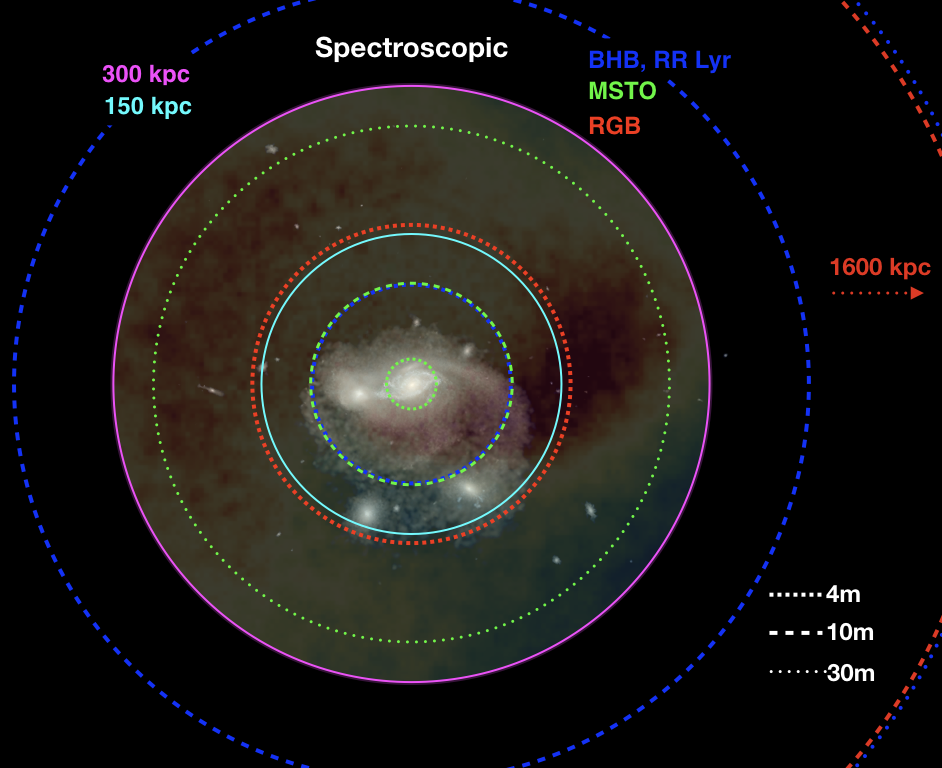} 
    \end{tabular}
        \vspace{-3mm}
    \caption{Distances to which the individual stellar tracers summarized in Table \ref{tbl:tracers} will be detected by astrometric (left), and spectroscopic (right) instruments proposed for the next decade, superimposed on a simulated pair of galaxies resembling the MW and M31\cite{2018arXiv180604143G}. The panels have different scales; Galactocentric radii of 150 kpc (cyan; the limit of current tracer populations) and 300 kpc (magenta; the approximate virial radius of the MW) are shown for reference. In the right panel, a prediction for the density fluctuations in the MW's DM halo induced by the MW's interaction with the LMC\cite{Garavito-Camargo19} is superimposed; the LMC's present-day position is approximated by the bright satellite to the left of the central MW-like galaxy.}
    \vspace{-5mm}
    \label{fig:distances}
\end{figure*}

\vspace{-3mm}
The Milky Way (MW) is a unique laboratory to study two fundamental questions in astrophysics:\\
\noindent\fbox{%
    \parbox{\textwidth}{{
    \vspace{1mm}
    \bf
\ding{228} How do galaxies of different masses form stars and evolve over cosmic time?\\
\ding{228} What is dark matter?\\
\vspace{-3mm}
}}}
Ours is the only galaxy for which we can hope to obtain both {\bf complete three-dimensional positions and velocities} (6D phase-space positions) and {\bf multiple elemental abundance measurements} for individual stars across the Hertzsprung-Russell diagram and throughout the Galaxy to its virial radius (Figure \ref{fig:distances}). This multidimensional view is key to answering these two questions.

The halo of the MW contains satellite galaxies, globular clusters (GCs), disk stars ejected via outflows or mergers\cite{villalobos08,purcell10, laporte18b}, and remnants of tidally disrupted dwarf galaxies and star clusters in the form of stellar streams\cite{GrillmairCarlin:2016}. Together, these components comprise {\bf a treasure trove of information about the formation of our Galaxy and its companions.} Because of the long dynamical times ($\sim$1--3 Gyr) in the halo, the 6D phase space properties of even individual MW halo stars retain imprints of their origin and therefore the MW's accretion history, while their formation environment is recorded in their chemical abundances. We can therefore use a multidimensional view of the halo to {\bf reconstruct the different accretion events} that contributed to the MW's mass assembly and compare these to existing satellite galaxies. The multidimensional stellar halo thus also contains clues to how dwarf galaxies form on even the smallest mass scales over cosmic time. 

{\bf Stellar streams} provide a snapshot of Milky Way stellar halo formation in action and improve our understanding of its fundamental building blocks. Wide-area imaging surveys have discovered more than 50 stellar streams\cite{GrillmairCarlin:2016}, with more expected from Large Synoptic Survey Telescope (LSST), but full 6D phase-space information exists for less than 20\%.
Obtaining this information, combined with detailed metallicity and chemical abundance measurements of stellar streams, will provide us the ages, formation times, and orbits of their progenitors, helping bridge the gap between the smooth stellar halo and the present-day MW satellite population. Using spectroscopy to complete the phase-space picture and add abundances for individual halo stars in the field is particularly crucial to identify the oldest, fully phase-mixed streams\cite{2018AJ....156..179R} (see also the whitepaper ``Local Dwarf Galaxy Archaeology'' by A. Ji et al.). Individual orbits of dwarf galaxies and GCs based on accurate 3D motions---obtained by averaging 10s to 1000s of stars---inform us about the formation and dynamical evolution (e.g. by tidal perturbation) of dwarf galaxies and globular clusters. 

The halo starts at the {\bf disk-halo interface}, which holds clues to the accretion history of the MW in the time- and mass-dependent imprints of interactions between the disk and satellite galaxies\cite{minchev09,gomez12}. So far, only a small fraction of this region has been explored: 6D phase-space alone cannot separate the origin of the stellar populations, but must be combined with chemical abundances and ages to reconstruct the accretion history and probe the flattening of the Galaxy in the vertical direction \cite{price-whelan15,bergemann18,dimatteo18}. Sensitivity to different timescales also offers the unique opportunity to constrain the orbital mass-loss history of the Sagittarius (Sgr) dwarf galaxy \cite{laporte18b}: the inner disk is sensitive to the most recent pericentric passage\cite{laporte18e}, while in the outer disk ($R\sim 16\,\rm{kpc}$) signs of incomplete phase-mixing due to prior pericentric passages may still be visible \cite{laporte18e}. Phase-space maps of these structures\cite{antoja18} can constrain the Galactic potential in the midplane \cite{darling19} and test whether the MW has a dark disk \cite{read08}, but studies with Gaia DR2 reach only to $R_{\rm{GC}}\sim10\,\rm{kpc}$ \cite{laporte18e}. 

The visible components of the halo are also crucial for constraining the total mass, profile and shape of {\bf the MW's dark matter (DM) halo}. The {\bf mass} of the Milky Way's halo determines which simulated galaxies we should compare to when assessing the consistency of our observations with predictions sensitive to the DM model, such as the number and structure of satellite galaxies. The concentration and {\bf radial profile} of the Galaxy's DM halo set constraints on its accretion history and formation time\cite{2002ApJ...568...52W}, which are responsible for some of the remaining scatter in comparisons with simulations\cite{2015ApJ...810...21M}. Measuring 3D motions of dynamical tracers in the halo is required to constrain both the mass profile and the total MW mass; our current understanding is confined to $< 40$~kpc from the Galactic center. The {\bf shape} of the Galactic halo could potentially differentiate between different DM models \cite{2018MNRAS.479..359S,2018PhR...730....1T,2015PhRvD..91b4022K} but constraints are presently inconclusive, hampered by insufficient data and overly simplistic models.

Stellar streams offer the unique capability to measure both the enclosed mass and the three-dimensional shape of the gravitational field \cite{1999ApJ...512L.109J}.\
%cite{Johnston:1998,Erkal:2016a,Bovy:2016}.
By coherently tracing the gravitational potential, stellar streams also provide enough information to constrain realistic models of the MW mass distribution, including contributions from halo triaxiality\cite{Law:2005,2010ApJ...714..229L,2012MNRAS.424L..16L,2015ApJ...799...28P} and from the MW disk and bar \cite{2016MNRAS.460..497H,2016ApJ...824..104P,2017NatAs...1..633P,Banik:2018}.
Recent studies show that full phase-space measurements of a dozen streams will be able to constrain the Milky Way halo parameters at the sub-percent level \cite{2015ApJ...801...98S,Bonaca:2018}.
Combined measurements of the 3D motions of dwarf galaxies, globular clusters, stellar streams, and individual halo stars, out to the edge of the MW, will allow us to {\bf constrain the MW mass sufficiently to relieve systematic uncertainties} in comparisons with simulations. Many proposed tests of dark matter in the Milky Way require constraints on the properties of its dark matter halo on scales comparable to the virial radius.  Since the region where we have good constraints is set by where we have data, {\bf we must observe stars to the MW's edge} or beyond in order to carry out these tests. These stars are sparsely distributed but not non-existent, and are likely to be in tidal streams that can cross out of the MW's sphere of influence \cite{2017MNRAS.470.5014S}. Indeed, Figure \ref{fig:distances} and Table \ref{tbl:tracers} illustrate that proposed capabilities offer the ability to identify and map distances and motions of individual stars {\bf beyond the viral radii of either the Milky Way or Andromeda, throughout the Local Group}. This opens a new, barely explored frontier in which to challenge our models of cosmological structure formation\cite{2006AJ....131.2980S,2009ApJ...707L..22T}.

{\bf The MW's interaction with the Large
Magellanic Cloud (LMC)} likewise provides a unique opportunity to test predictions
from dark matter models and refine estimates of the MW's mass distribution. Given its proximity, high mass, and recent orbit
through the stellar and DM halo \cite{Besla07, Besla12}, the interaction between the MW and the LMC induces time-dependence in the combined gravitational potential\citep{Weinberg95,gomez15} at distances larger than $\sim$15 kpc\citep{laporte18a}. Recent multidimensional measurements of the Orphan Stream\cite{Koposov18}, at the inner edge of this range, show tentative evidence of this effect\cite{Erkal18b}, but the strongest signal---and the strongest constraints on the time-dependence of the gravitational potential---is in the outer halo. Particle-based DM models also predict the existence of a DM ``wake'' from scattering and resonances between DM particles and the 
LMC's orbit \cite{White83, Weinberg89, Weinberg95}. The properties (morphology, amplitude, and kinematics) of the wake depend sensitively on the nature of the DM particle \cite{Furlanetto02}. The wake is predicted to produce a
unique signal in the phase space of the stellar halo at distances of 50-100 kpc \cite{Garavito-Camargo19} (Figure \ref{fig:distances}, lower right panel). The detection and characterization of the DM wake within the outer stellar halo will permit new tests of the nature of the DM particle and
constrain the total dark matter mass of both the MW and the LMC at infall.

\begin{table*}[!tbp]
\centering
\begin{tabular}{l||l||l|l|l||l|l|l}
\hline
& \multicolumn{6}{c}{Distance range (kpc)} \\
\hline
 & Photometry & \multicolumn{3}{c||}{Spectroscopy} & \multicolumn{3}{c}{Astrometry} \\
    \hline
    Tracer & LSST & 30m 	&  10m 	& 4m & WFIRST & LSST & Gaia \\
    \hline
    \hline
    RGB (brightest) & 10000 & 4000  &	1500 &  600 & 10000 & 4000 & 600\\
    BHB, RR Lyr    & 1600 & 600	 &   250 &	100 & 1600 & 600 & 250\\
    RGB (faintest)	 & 600  &  250	 &   100 &   40 & 600 & 250 & 40\\
    MSTO	         & 400 &  150	 &    60 &   25 & 400 & 150 & 25\\
    \hline
\end{tabular}
\vspace{-3mm}
\caption{Distance to various stellar tracers at the limiting magnitudes of surveys planned for the 2020s: photometric (the Large Synoptic Survey Telescope, LSST, assuming a limiting magnitude of $r\sim$ 26.5), spectroscopic (30-m, 10-m, and 4-m spectroscopic facilities, assuming a limiting magnitude of $r\sim$ 24.5, 22,5, and 20.5, respectively) and astrometric (the Wide-Field InfraRed Space Telescope (WFIRST) High-Latitude Survey, LSST, \& Gaia, assuming limiting magnitudes of $r\sim$  26.5, 24.5, and 20.5, respectively).}\label{tbl:tracers}
\vspace{-5mm}
\end{table*}

\vspace{-7mm}

\section*{Important tracers of structure}
\vspace{-4mm}
Observing stars across the HR diagram offers the chance to optimize between number density, accurate distance measurements, and distance range in studying the MW halo. {\bf Main-sequence (MS) stars} are present in all stellar populations, regardless of age or metallicity. Their high number densities will enable the discovery of structures in the halo that cannot be resolved by giants alone. For example, accessing the MS population is essential for identifying and studying the smallest progenitor galaxies in the MW stellar halo, since ultra-faint dwarfs (UFDs), which probe how galaxies populate DM halos at the threshold of galaxy formation, contain only a handful of evolved stars. {\bf Blue horizontal branch} (BHB) and {\bf red clump} (RC) stars offer larger range but with a higher degree of contamination in the field. For BHBs this problem can be alleviated with u or b band imaging; RCs become a robust distance indicator once a structure at common distance is identified. These two tracers offer complementarity in terms of color and range: BHBs are brighter in bluer bands while RCs are brighter in the red. Finally, {\bf RR Lyrae} stars are perhaps the most useful standard candle for halo studies, although biased toward low-metallicity populations. They provide accurate distances (2\%) with no foreground/background contamination. Upcoming time-domain surveys like LSST have the depth to {\bf discover every RR Lyrae star within $R_{\mathrm{vir}}$}. Spectroscopic follow-up is the main challenge for these tracers: to obtain RVs near $R_{\mathrm{vir}}$ will require reaching $g \simeq$ 22--23 in an integration time of $\lesssim 15$ minutes, given their typical pulsation period of 6-12 hours. Combined analysis of different tracers can maximize our multidimensional view of the halo by transferring well-measured quantities from one stellar population to another within the same structure. Since these measurements will likely come from many different instruments, {\bf efficient, flexible, open-source} strategies for cross-matching, joint analysis, and synthetic observations  will be crucial ( see the whitepaper by M. Ness et al, ``In Pursuit of Galactic Archaeology'').
%\subsubsection*{The importance of distance measurements}
 %- RRLe - unambiguous distances/IDs, no  contam, hard to get spx\\

%\subsubsection*{Other useful tracer stellar populations with special considerations}

% - BHBs - bright in bluer, less bright in redder bands; contamination problem in the field w/o  (blue stragglers). At a well defined distance not an issue\\
% - RC - brighter at redder bands,  (but not in field)\\

\vspace{-7mm}

\section*{Resources Needed: Theory and Simulation}
\vspace{-4mm}
%[Summary table]
%\noindent \textbf{\textit{}}:
Characterizing the formation and mass distribution of the MW requires a wide range of theoretical and numerical tools working in tandem. (1) {\bf Ab-initio cosmological hydrodynamical zoom-in simulations} offer us a way to study the essential physics of the formation of MW-like galaxies in cosmological context, and connect the MW and its satellites with galaxy populations at both high and low redshifts. (2) {\bf Controlled numerical experiments} via N-body methods treat detailed dynamical questions pertaining specifically to the Milky Way and its environment. (3) {\bf Development of non-equilibrium models} for the MW will account for the time-dependence of the MW through basis function expansion methods. (4) {\bf Synthetic observations} of simulations\cite{2018MNRAS.481.1726G,2018PASP..130g4101R,2018arXiv180610564S} allow us to test sophisticated analysis methods on realistic data sets with known properties and train machine-learning algorithms. Some of these approaches require major allocations on state-of-the-art computing centers, while others are well matched to second-tier resources with lower subscription rates. {\bf Maintaining at least some clusters after upgrades to new systems} would help, as would {\bf continued funding for theoretical work} on advanced MW models.

\vspace{-7mm}

\section*{Resources Needed: Observation}
\vspace{-4mm}
To achieve the science goals outlined in Section 3 will require instruments and observational campaigns tailored to the {\bf faint, low-density} structures containing most of the information in the halo. Typical target densities in a stellar stream, for example, can be as low as one per square degree. An ideal instrument will have a {\bf wide field of view} (FOV; on the order of $\sim$1 deg$^2$) and sufficient aperture to {\bf reach the MSTO at the virial radius} (300 kpc; a distance modulus of $\sim$20). Because cosmological accretion is stochastic and asymmetric, obtaining a complete accretion history of the MW will require {\bf broad coverage at high Galactic latitudes in both hemispheres}.

\vspace{2mm}
\noindent \textbf{\textit{Astrometry}}:
\textit{Gaia} will measure PMs for over 1 billion stars; however, because of its limiting magnitude of $G\sim 20$, the most distant stars in the \textit{Gaia} sample will be at 100 kpc, only $\sim1/3$ of the way to the virial radius (Figure \ref{fig:distances}). In addition, beyond $\sim$ 15 kpc in the halo, \textit{Gaia} (and subsequent spectroscopic follow-up programs) cannot provide kinematic information for individual main sequence stars. Conversely, multi-epoch \textit{HST} imaging has been exploited to make extremely accurate PM measurements of resolved stellar systems in the Local Group (LG) \cite{Sohn2012a, Kallivayalil13, Sohn18}, as well as individual MW halo stars \cite{Deason2013b,Sohn16,Cunningham2018b}. However, given the small FOV of \textit{HST}, these samples are limited to deep pencil beams. LSST will provide phenomenal photometric depth over the southern sky, and return PMs for stars in the magnitude range $r \sim$ 19--24, but the PM measurement uncertainties are too large to study individual stars at large distances. At $R_{Vir} \sim 300$ kpc, projected LSST PM uncertainties ($\sim0.5$ mas/yr at $r\simeq23$) correspond to a velocity uncertainty of $\sim 700$ km/s, while the circular speed is $\sim$150 km/s.
The Wide Field Imager (WFI) planned for WFIRST will provide the ideal complement to LSST. With 100 times the FOV area and similar image quality to \textit{HST}, {\bf WFIRST will achieve both the precision and depth of astrometry needed to target the faint MW stellar halo} with measurement uncertainties as low as 25 $\mu$as yr${}^{-1}$ in the High Latitude Survey\cite{2017arXiv171205420T} of about 2200 deg$^2$ (at high Galactic latitudes ideal for studying the outer halo). The Euclid mission will cover more area, but its pixel scale is too large to yield good astrometry.

\vspace{2mm}

\noindent\textbf{\textit{Spectroscopy}}:
Spectroscopic surveys with 4m class telescopes (e.g., WEAVE, DESI, 4MOST) expected to start in 2020-2021 have limiting magnitudes aligned with the \Gaia\ depth ($G\sim20-21$). To go deeper, a larger aperture is needed.  Most existing 8-10m telescopes are not capable of doing the science proposed here due to their small FOV, usually $20'$ in diameter or less; in contrast, DESI (on the 4m Mayall telescope) has a $\sim3^\circ$ FOV.
The only funded project with sufficient FOV for stream studies is the Prime Focus Spectrograph on Subaru. Another option is the Maunakea Spectroscopic Explorer, a currently unfunded 11.25m telescope with a FOV of $1.4^\circ$ in diameter. {\it Neither} of these currently planned projects allows access to the general US astrophysics community. {\bf In order to achieve the science described above, the US community must have access to one of the 10m class, wide FOV telescopes capable of conducting a survey of the MW's outer halo}. The two planned (but not fully funded) extremely large telescopes (30m class) planned in the U.S., the Giant Magellan Telescope and the Thirty Meter Telescope, will reach 1-2 magnitude deeper than the 10m class. Due to their small FOV ($<$20'), \textbf{multiplexing and multithreading} (i.e. opportunistic scheduling of individual spare fibers) will be a crucial capability to reach the edge of the MW and beyond.

%The extremely large telescopes (ELTs, 30m class) will be able to go another 1-2 magnitude deeper than 10m class. The two planned (but not fully funded) extremely large telescopes (30m class) planned in the U.S., the Giant Magellan Telescope and the Thirty Meter Telescope, will reach 1-2 magnitude deeper than the 10m class. Due to their small FOV ($<20'$), a full sky survey would be difficult, but a small patch of the sky with deep observations is still possible.

%\FIMEX{should we say something here like "in order to achieve the proposed science described above, it is important to make sure that US community will have access to one of the 10m class, wide FOV telescope and conduct a Milky Way survey."}

\newpage

\bibliographystyle{JHEP}
\bibliography{references}

\clearpage
\subsection*{Affiliations}
\input{affiliations}

\end{document}

%% file: authors.tex
\def\altaffilmark#1{\textsuperscript{#1}}
\def\affil#1{\noindent #1 \\}

\noindent {\bf Co-authors (affiliations after text):}
\begin{raggedright}
\small
Jeffrey~L.~Carlin\altaffilmark{1},
Emily~C.~Cunningham\altaffilmark{2},
Nicolas~Garavito-Camargo\altaffilmark{3},
Puragra~Guhathakurta\altaffilmark{2},
Kathryn~V.~Johnston\altaffilmark{4,5},
Chervin~F.~P.~Laporte\altaffilmark{6},
Ting~S.~Li\altaffilmark{7,8},
S.~Tony~Sohn\altaffilmark{9}
\end{raggedright}

\vspace{3mm}
\noindent {\bf Endorsers (affiliations after text):}\\
\begin{raggedright}
\small
{\it From the WFIRST Astrometry Working Group:}\\
Jay Anderson\altaffilmark{9},
Andrea Bellini\altaffilmark{9},
David P. Bennett\altaffilmark{10},
Stefano~Casertano\altaffilmark{9},
S.~Michael~Fall\altaffilmark{9},
Mattia Libralato\altaffilmark{9},
Sangeeta~Malhotra\altaffilmark{10},
Leonidas~A.~Moustakas\altaffilmark{11},
Jason Rhodes\altaffilmark{11}\\

{\it Other Endorsers:}\\
Lee~Armus\altaffilmark{12},
Yumi Choi\altaffilmark{13,14},
Andres del Pino\altaffilmark{9},
Elena~D'Onghia\altaffilmark{15},
Mark Fardal\altaffilmark{9},
Karoline~M.~Gilbert\altaffilmark{9},
Carl~J.~Grillmair\altaffilmark{12},
Nitya Kallivayalil\altaffilmark{16},
Evan~N.~Kirby\altaffilmark{17},
Jing~Li\altaffilmark{18},
Jennifer~L.~Marshall\altaffilmark{19},
Adrian~M.~Price-Whelan\altaffilmark{20},
Elena Sacchi\altaffilmark{9},
David~N.~Spergel\altaffilmark{5,20},
Monica~Valluri\altaffilmark{21},
Roeland P.~van der Marel\altaffilmark{9}
\end{raggedright}

%% file: affiliations.tex
\affil{$^{1}$ Large Synoptic Survey Telescope, 950 North Cherry Ave, Tucson, AZ 85719}
\affil{$^{2}$ Department of Astronomy \& Astrophysics, University of California Santa Cruz, 1156 High St, Santa Cruz, CA 95064, USA}
\affil{$^{3}$ Steward Observatory, University of Arizona, 933 North Cherry Avenue,Tucson, AZ 85721, USA}
\affil{$^{4}$ Department of Astronomy, Columbia University, 550 West 120th Street, New york, NY 10027}
\affil{$^{5}$ Center for Computational Astrophysics, Flatiron Institute, 162 5th Ave, New York, NY 10010, USA}
\affil{$^{6}$ CITA National Fellow, Department of Physics \& Astronomy, University of Victoria, 3800 Finnerty Road, Victoria BC, Canada V8P 5C2}
\affil{$^{7}$ Fermi National Accelerator Laboratory}
\affil{$^{8}$ Kavli Institute of Cosmological Physics, University of Chicago}
\affil{$^{9}$ Space Telescope Science Institute, 3700 San Martin Dr, Baltimore, MD 21218, USA}
\affil{$^{10}$ Goddard Space Flight Center, NASA.}
\affil{$^{11}$ Jet Propulsion Laboratory, California Institute of Technology, Pasadena, CA 91109}
\affil{$^{12}$ California Institute of Technology/IPAC}
\affil{$^{13}$ Department of Physics, Montana State University, P.O. Box 173840, Bozeman, MT 59717, USA}
\affil{$^{14}$ Steward Observatory, University of Arizona, 933 North Cherry Avenue, Tucson, AZ 85721, USA}
\affil{$^{15}$ Department of Astronomy, University of Wisconsin, Madison, 475 N Charter Street, Madison, 53706 WI}
\affil{$^{16}$ Department of Astronomy, University of Virginia, 530 McCormick Rd, Charlottesville VA 22904, USA}
\affil{$^{17}$ California Institute of Technology}
\affil{$^{18}$ Physics and Space Science College, China West Normal University, 1 ShiDa Road, Nanchong 637002, China}
\affil{$^{19}$ George P. and Cynthia Woods Mitchell Institute for Fundamental Physics and Astronomy, and Department of Physics and Astronomy, Texas A\&M University, College Station, TX 77843,  USA}
\affil{$^{20}$ Department of Astrophysical Sciences, Princeton University, 4 Ivy Lane, Princeton, NJ 08544, USA}
\affil{$^{21}$ Department of Astronomy, University of Michigan, 311 West Hall, 1085. S. University Ave. Ann Arbor MI 48109, USA}